# qLOOK: A Minimal Information System for Digital Storage and Reproducible Analysis of qPCR experiments

*Mirco Castoldi*

*Department of Gastroenterology, Hepatology and Infectious Diseases, Medical Faculty and University Hospital, Heinrich Heine University Düsseldorf, Germany*

*Center for Integrated Oncology Aachen Bonn Cologne Düsseldorf (CIO ABCD), Düsseldorf, Germany*

**Corresponding author:**

Mirco Castoldi, <mirco.castoldi@hhu.de>

**Word count**: *1846*

**Number of figures and tables**: *3*

**Financial support:** *This work was supported by a grant of the German Research Foundation (DFG, Deutsche Forschungsgemeinschaft) to M.C. (Reference: CA 830/3-1)*

**Conflicts of interest:** *The author has no relevant financial or non-financial interests to disclose*

**Abstract**

**Objective**: *Quantitative real-time PCR is widely used for gene expression analysis, yet inconsistencies in data storage and reporting limit reproducibility. While MIQE guidelines define the minimal information required for publication, they do not specify structured digital storage formats compatible with long-term reanalysis. This work presents qLOOK (qPCR-LOg-boOK), a tool for standardized digital storage and reproducible analysis of qPCR experiments.*

**Results**: *qLOOK is a modular R-based system that extracts data from Thermo Fisher/ABI .EDS files, formats it into a structured table (qLOOK_Data.xlsx), performs normalization and statistical analysis, and generates a log file (qLOOK_Summary.txt) recording reference genes, calibrators, and analytical parameters. All required R libraries are automatically installed and loaded, allowing researchers without coding experience to use the scripts. By preserving the qLOOK_Data table and the qLOOK_Summary log, users can reproduce or extend analyses without reprocessing raw files. While currently limited to .EDS files, the modular design allows adaptation to additional qPCR formats in the future.*

*Besides providing an easy and transparent approach to analyze qPCR experiments, qLOOK also provides a minimal, standardized, and transparent solution for digital documentation, enhancing reproducibility, supporting long-term data stewardship, and facilitating integration into electronic laboratory notebooks or publication supplementary material.*

**Introduction**

*Quantitative real-time PCR (qPCR) is a cornerstone technique in molecular biology, enabling precise quantification of gene expression and nucleic acid abundance across diverse biological applications. Despite its widespread use, qPCR remains one of the most variable and inconsistently reported experimental methods. To improve transparency and reproducibility, the Minimum Information for Publication of Quantitative Real-Time PCR Experiments (MIQE) guidelines were introduced in 2009 (1, 2), providing standards for experimental design, data analysis, and reporting. While MIQE defines what information should be reported, it does not specify how these data should be digitally stored and preserved. In practice, qPCR data are often kept as instrument-specific output files, accompanied by manually curated spreadsheets or text documents. This unstructured approach poses several challenges: essential metadata are frequently missing or inconsistently recorded; raw data may only be accessible through proprietary software; and analytical steps such as normalization or calibrator are rarely documented in a reproducible way. Consequently, even within the same laboratory, reproducing or reanalyzing experiments conducted years earlier can be difficult.*

*The adoption of electronic laboratory notebooks (ELNs) and data management mandates has increased the need for standardized digital storage. ELNs provide a framework for traceable data capture but currently lack templates for qPCR data organization and processing. To address this gap, this work introduces qLOOK (qPCR-LOg-boOK), a minimal information system for digital storage, reproducible analysis, and reporting of qPCR experiments. qLOOK translates MIQE principles into a practical, script-based digital format compatible with ELNs and open-source environments. qLOOK consists of three modular R scripts that (i) extract essential experimental data from raw .EDS files (Thermo Fisher/ABI), (ii) perform normalization and statistical analysis, and (iii) generate visualizations and summaries. The core outputs are two structured text files: qLOOK_Data, containing formatted tabular qPCR data, and qLOOK_Summary, a log of all analytical steps. Together, these files form a reproducible,*

*platform-independent data package, allowing any user with access to R and the scripts to reproduce or extend the analysis, independent of proprietary software or hardware.*

*Beyond technical reproducibility, qLOOK supports open and digital research workflows. Including qLOOK outputs in ELNs or as supplementary material enhances transparency, facilitates data quality assessment, and aligns with FAIR (3) (Findable, Accessible, Interoperable, and Reusable) principles. The system is intentionally minimal, focusing on essential elements for reproducible storage and reanalysis. While currently supporting Thermo Fisher .EDS files, its structure can be readily adapted to additional formats, enabling broader standardization of qPCR data in digital research environments.*

**Design principles and scopae**

*qLOOK was designed to provide a minimal, reproducible, and broadly accessible system for analysis of qPCR experiments and digital storage. Its primary goal is to allow researchers to preserve essential experimental data and analytical steps in a structured format that enables full reanalysis, independent of proprietary software or instrument-specific tools. While compatibility with ELNs and FAIR data principles emerged as a beneficial feature, the core design focuses on simplicity, portability, and reproducibility.*

*The script works with EDS file generated by Thermo Fisher/ABI software and has been successfully tested on multiple cyclers, including 7500, 7500Fast, StepOnePlus, Viia7, QuantStudio6 and QuantStudio3. R was selected as the implementation language to ensure platform independence and accessibility. The scripts are fully modular and require no coding experience; user input is handled via pop-up windows, and a future option for self-contained Windows executables is envisioned for users who do not wish to install R or RStudio, or working in environments with restricted software installation.*

*A key guiding principle of qLOOK is to keep metadata minimal but sufficient for complete reanalysis of experiments. This approach aligns with MIQE guidelines and FAIR data practices,*

*ensuring that stored qPCR data are transparent, reusable, and reproducible. By balancing simplicity with rigorous documentation, qLOOK provides a standardized framework suitable for individual laboratories, collaborative projects, and broader academic sharing.*

**Implementation and workflow**

*qLOOK is implemented as a modular, script-based workflow in R, consisting of three independent scripts—qLOOK_Module1_v1.0.R, qLOOK_Module2_v1.0.R, and qLOOK_Module3_v1.0.R. The modular architecture was intentionally chosen to allow users to execute only the relevant parts of the pipeline, such as re-running normalization or statistical analyses without repeating data extraction. Each module operates through graphical pop-up windows, requiring no coding skills or prior R experience (**Figure 1**).*

*qLOOK_Module1_v1.0.R performs data extraction and formatting. Upon selecting a folder containing .EDS files, the script automatically identifies and processes all available files—independently of cycler model—and compiles the results into a structured spreadsheet, qLOOK_Data.xlsx, containing formatted Cq values and metadata. It also generates a reference gene stability report (qLOOK_RefGenes.xlsx) using ΔCq (4), GeNorm (5), and NormFinder (6) algorithms, and a log file (qLOOK_Summary.txt) that records all processing steps. Importantly, qLOOK_Module1_v1.0.R's data-parsing logic is modular by design. This allows the program to automatically recognize and process EDS archives with different internal structures. These include those generated by older software versions, such as 7500, 7500 Fast, StepOnePlus, Viia7, and QuantStudio6, as well as the newer Thermo Fisher DA2 real-time PCR software for QuantStudio3 and QuantStudio5 cyclers.*

*qLOOK_Module2_v1.0.R performs data normalization. After the user selects the qLOOK_Data.xlsx file, the script prompts for reference genes and a calibrator sample. The resulting files include qLOOK_Norm.xlsx (normalized and expression data) and qLOOK_Express.xlsx, which provides three linked sheets containing normalized values,* *relative expression in linear format (ΔΔCq), and using the Livak method (2^-ΔΔCq; (7)). This module also generates distribution plots of 2^-ΔΔCq values and records all analytical steps, reference genes, and calibrator information in qLOOK_Summary.txt.*

*qLOOK_Module3_v1.0.R handles statistical analysis and data formatting for downstream applications. It imports qLOOK_Express.xlsx, requests a p-value threshold (default = 0.05), and performs one-way ANOVA and pairwise t-tests analyses. The results are stored in qLOOK_ANOVA.xlsx and qLOOK_TTEST.xlsx, including both unfiltered and significance-filtered comparisons. Additional outputs include a GraphPad-compatible file (qLOOK_GraphPad.xlsx) and a folder of automatically generated box plots (qLOOK_Plots). All computational steps are appended to the qLOOK_Summary.txt log, ensuring complete reproducibility.*

**Data structure and metadata model**

*The qLOOK system is designed around a minimal yet sufficient data model, ensuring that the essential elements required for qPCR data reanalysis are preserved in a standardized and transparent format. The output of the workflow centers on two primary files: qLOOK_Data.xlsx and qLOOK_Summary.txt.*

*qLOOK_Data.xlsx (**Table 1**) represents the structured data table generated from processed .EDS files. The first column lists sample identifiers, while each subsequent column corresponds to a target gene, with numerical entries representing Cq values. This matrix-style format allows straightforward visualization, reanalysis, and compatibility with standard statistical or plotting tools. Because the script can process large batches of .EDS files, users are advised to maintain consistent sample naming conventions and avoid spaces in names to ensure error-free handling in R. The table does not currently include experimental metadata (e.g., operator, instrument, or plate ID), as these parameters are already defined within MIQE* *guidelines and are intended to be stored separately in laboratory documentation. However, the structure is readily extensible to accommodate future metadata layers if required.*

*qLOOK_Summary.txt (**Table 2**) functions as a comprehensive analytical log that records all key parameters influencing data interpretation. It includes script versions, timestamps, lists of processed .EDS files, instrument type, reference gene(s), calibrator sample(s), selected statistical thresholds, list of input files, and the names of all generated output files. Each computational step is chronologically documented in plain text, providing full transparency and reproducibility of the analytical workflow.*

*Together, qLOOK_Data.xlsx and qLOOK_Summary.txt constitute the minimal dataset required for complete reanalysis of qPCR experiments, in line with FAIR data principles. This format complements, rather than replaces, MIQE standards—translating them into a practical digital framework for ELN integration and long-term data stewardship*

**Reproducibility and interoperability**

*Reproducibility was a primary consideration in the design of qLOOK. Each analytical step is automatically recorded in qLOOK_Summary.txt, providing a transparent record of data extraction, normalization, and statistical analysis. The log includes all relevant parameters, such as script versions, reference genes, calibrators, and statistical thresholds, allowing users to reproduce identical results or apply alternative analytical settings using the same input data. Because all operations are performed through scripted procedures rather than manual data handling, qLOOK minimizes user-dependent variation and ensures consistency across analyses.*

*The separation between raw instrument output (.EDS files) and the standardized qLOOK_Data.xlsx format further supports reproducibility and long-term accessibility. Once extracted, data can be reanalyzed independently of the original proprietary software, reducing dependency on specific software versions or operating systems. By relying solely on open-* *source R functions and generating plain-text or spreadsheet outputs, qLOOK maintains full transparency and platform independence.*

*Regarding interoperability, qLOOK was designed to align with FAIR and MIQE principles. The compact metadata structure allows straightforward incorporation into ELNs and compatibility with institutional or public data repositories. The modular code base also permits the inclusion of additional qPCR file formats or metadata fields without altering the core structure. Together, these features make qLOOK a practical and extensible tool for standardized digital storage and reproducible reanalysis of qPCR experiments.*

**Discussion and Conclusion**

*qLOOK provides a simple and standardized approach for storing, processing, and reanalyzing qPCR data independently of proprietary software. By retaining qLOOK_Data.xlsx and qLOOK_Summary.txt, users preserve the minimal information required to reproduce or extend analyses without reprocessing the original .EDS files, ensuring transparent data provenance and enabling future reanalysis. The current implementation requires R, RStudio, and RTools, but all necessary libraries are automatically installed and loaded, facilitating use by researchers without prior coding experience. A self-contained, R-independent Windows executable is under development to increase accessibility in environments with restricted administrative privileges. Although the current version focuses on Thermo Fisher/ABI .EDS files, the modular design of the scripts allows straightforward inclusion of additional file formats from other qPCR cyclers in the future. Overall, qLOOK represents a practical tool for minimal, standardized digital documentation of qPCR experiments, supporting reproducibility, transparency, and long-term data stewardship in molecular biology and biomedical research.*

**qLOOK Accessibility**

*The qLOOK scripts, ReadMe file, and license information are available for download from M.C.'s GitHub repository: [https://github.com/mircocastoldi](https://github.com/mircocastoldi)*

**Abbreviations**

*qLOOK; qPCR-LOg-boOK, MIQE; Minimum Information for Publication of Quantitative Real-Time PCR Experiments. qPCR; Quantitative real-time PCR. FAIR; Findable, Accessible, Interoperable, and Reusable, ELN; electronic laboratory notebooks.*

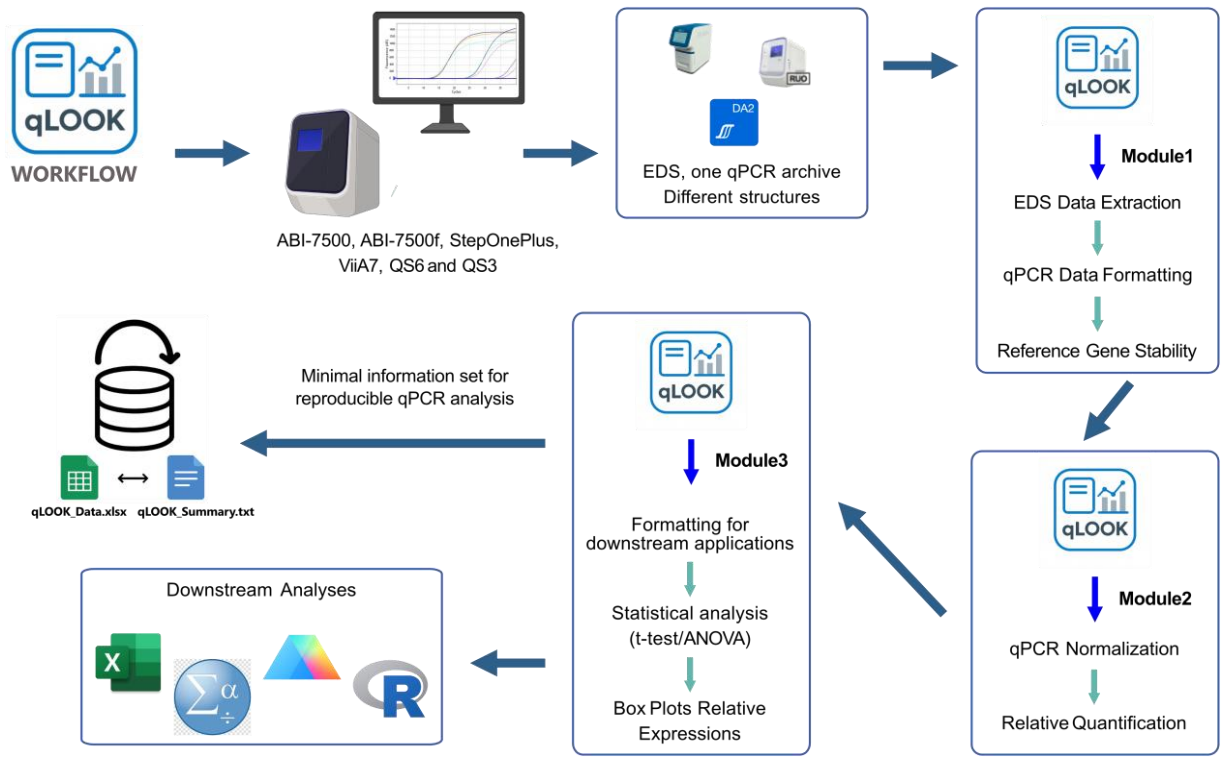

**Figure 1** *shows the schematic representation of the qLOOK Workflow.*

| Sample_Name | ANLN | AURKb | GAPDH | KIF11 |
|---|---|---|---|---|
| phx_153_4h | 22,3512 | 29,1157 | 20,6971 | 20,5673 |
| phx_162_8h | 21,0680 | 21,9792 | 21,4744 | 20,5262 |
| phx_163_4h | 22,8379 | 25,1766 | 20,6408 | 20,3407 |
| phx_170_8h | 21,8604 | 24,5755 | 21,1083 | 19,3488 |
| phx_174_48h | 14,9019 | 17,5523 | 19,8234 | 16,1102 |
| phx_176_48h | 14,6684 | 17,5083 | 18,9228 | 15,7772 |
| phx_190_8h | 21,2842 | 23,7515 | 19,9763 | 19,7595 |
| phx_194_48h | 14,4002 | 16,8717 | 19,1346 | 15,4296 |
| phx_195_48h | 15,2774 | 18,0843 | 20,1057 | 16,5263 |
| phx_22_48h | 15,9776 | 17,8521 | 20,8105 | 16,5250 |
| phx_26_24h | 16,5454 | 18,4588 | 20,7834 | 18,3250 |
| phx_28_24h | 18,4952 | 20,7846 | 20,6637 | 19,6882 |
| phx_31_24h | 17,5852 | 19,7550 | 19,4581 | 18,2496 |
| phx_52_24h | 21,3693 | 22,8459 | 20,3576 | 19,4385 |
| phx_63_8h | 19,9575 | 23,0572 | 21,2054 | 19,2469 |
| phx_65_4h | 21,4997 | 25,9957 | 21,9071 | 19,2268 |
| phx_70_8h | 25,2151 | 23,0949 | 20,5137 | 19,6649 |
| phx_71_8h | 20,7688 | 22,5058 | 20,2955 | 19,4323 |
| phx_72_4h | 21,8862 | 23,8683 | 21,4390 | 21,7104 |
| phx_73_4h | 23,0833 | 25,6393 | 21,7183 | 19,6814 |
| sham_160_8h | 34,2735 | 25,6100 | 21,2632 | 20,1316 |
| sham_161_4h | 21,6394 | 25,0499 | 20,5429 | 19,0115 |
| sham_169_24h | 22,9606 | 23,4582 | 20,2661 | 19,7673 |
| sham_171_24h | 20,8130 | 23,6791 | 20,7537 | 19,8000 |
| sham_172_4h | 21,1039 | 23,2030 | 20,1256 | 18,3395 |
| sham_178_8h | 23,6461 | 24,3788 | 21,8180 | 20,5194 |
| sham_180_4h | 20,9927 | 24,2251 | 20,9408 | 19,1501 |
| sham_185_48h | 20,9911 | 23,1441 | 21,6079 | 19,1118 |
| sham_64_8h | 21,6082 | 23,5504 | 20,3015 | 19,7384 |
| sham_66_48h | 20,3126 | 21,9276 | 21,2515 | 18,6223 |
| sham_67_24h | 23,6023 | 24,3852 | 20,2212 | 19,7568 |
| sham_78_48h | 20,4342 | 21,9305 | 20,8713 | 18,1395 |
| sham_79_48h | 20,5357 | 22,9319 | 20,1377 | 18,8086 |
| sham_82_8h | 22,1111 | 26,3318 | 20,8869 | 20,4121 |
| sham_83_8h | 22,1228 | 23,6353 | 20,2361 | 20,6040 |
| sham_84_4h | 21,4856 | 30,6392 | 21,3718 | 19,9124 |
| sham_86_48h | 19,4260 | 21,5193 | 20,2765 | 17,8837 |
| sham_89_24h | 22,1744 | 23,5963 | 19,8232 | 19,8377 |
| sham_90_24h | 19,8258 | 21,8817 | 20,3324 | 19,0025 |

**Table 1** *shows the content of the qLOOK_Data.xlsx file after being generated by the qLOOK_Module1 script.*

```
 1  ----- qLOOK - qPCR-LOg-boOK -----
 2  ----- EDS Processing Log -----
 3
 4  --- 2025-10-15 13:42:21 ---
 5  Script: qLOOK_Module1_v1.0.R (v1.0)
 6  Script: qLOOK_RefGene_v1.R (v1)
 7  Original .EDS files processed:
 8  - ANLN_05092025.eds processed as TXT (analysis_result.txt)
 9  - AURKb_08092025.eds processed as TXT (analysis_result.txt)
10  - GAPDH_2025-09-10.eds processed as TXT (analysis_result.txt)
11  - KIF11_03092025.eds processed as TXT (analysis_result.txt)
12
13  If .TXT -> EDS was generated by ViiA7, StepOneplus, QuantStudio6, ABI7500, or ABI7500fast
14  If .JSON -> EDS was generated by QuantStudio3
15
16  Created files:
17  - qLOOK_qBASE_20251015_134207.xlsx
18  - qLOOK_qBASE_20251015_134207.txt
19  - qLOOK_Data_20251015_134207.xlsx
20  - qLOOK_RefGene_20251015_134207.xlsx
21
22
23  ----- qPCR Normalization and Relative Expression Log -----
24
25  --- 2025-10-15 13:44:00 ---
26  Script: qLOOK_Module2_v1.0.R (v1.0 (14-10-2025))
27  Input File: qLOOK_Data_20251015_134207.xlsx
28  Reference Gene(s): GAPDH
29  Calibrator Sample(s): sham_161_4h, sham_180_4h, sham_84_4h
30  Created Files:
31     - qLOOK_Norm_20251015_134236.xlsx
32     - qLOOK_Express_20251015_134236.xlsx
33     - Boxplots Folder: qLOOK_Norm_Plots_20251015_134353
34
35
36  ----- Statistical Analysis Log -----
37
38  --- 2025-10-15 13:46:14 ---
39  Script: qLOOK_Module3_v1.0.R (v1.0 (14-10-2025))
40  Input Excel: qLOOK_Express_20251015_134236.xlsx
41  p-value cutoff: 0.05
42  Created files:
43     - qLOOK_GraphPad_20251015_134607.xlsx
44     - qLOOK_ANOVA_20251015_134607.xlsx
45     - qLOOK_TTEST_20251015_134607.xlsx
46  Plots folder: qLOOK_Plots_20251015_134607
47  Plots:
48     - ANLN_boxplot.png
49     - AURKb_boxplot.png
50     - GAPDH_boxplot.png
51     - KIF11_boxplot.png
52     - Heatmap_20251015_134607.png
53
54
```

**Table 2** *shows the structure and content of the qLOOK_Summary.txt log file after the three qLOOK modules were run to analyze an experiment with four EDS files. In addition to logging the date, script version, and names of the input and output files, the qLOOK scripts log the metadata necessary for reanalyzing the experiment. This metadata includes the selected reference gene (GAPDH), the calibrator samples (Sham_161_4h, Sham_180_4h, and Sham_84_4h), and the p-value cutoff used to filter the results of the pairwise t-test (0.05).*